\documentclass[sigconf]{acmart}
\AtBeginDocument{%
  }

\usepackage{amsmath}
\usepackage{multirow}
\usepackage{soul}
\usepackage{booktabs}
\usepackage{balance}

\copyrightyear{2025}
\acmYear{2025}
\setcopyright{acmlicensed}\acmConference[WWW Companion '25]{Companion Proceedings of the ACM Web Conference 2025}{April 28-May 2, 2025}{Sydney, NSW, Australia}
\acmBooktitle{Companion Proceedings of the ACM Web Conference 2025 (WWW Companion '25), April 28-May 2, 2025, Sydney, NSW, Australia}
\acmDOI{10.1145/3701716.3717535}
\acmISBN{979-8-4007-1331-6/2025/04}

\begin{document}

\title{Topo Goes Political: TDA-Based Controversy Detection in Imbalanced Reddit Political Data}

\author{Arvindh Arun}
\authornote{Both authors contributed equally to this research.}
\email{arvindh.arun@ki.uni-stuttgart.de}
\affiliation{%
  \institution{International Institute of Information Technology, Hyderabad}
  \institution{University of Stuttgart}
  \city{Stuttgart}
  \country{Germany}
}

\author{Karuna K Chandra}
\authornotemark[1]
\email{karunakchandra@gmail.com}
\affiliation{%
  \institution{International Institute of Information Technology, Hyderabad}
  \city{Hyderabad}
  \country{India}
}

\author{Akshit Sinha}
\email{akshit.sinha@students.iiit.ac.in}
\affiliation{%
  \institution{International Institute of Information Technology, Hyderabad}
  \city{Hyderabad}
  \country{India}
}

\author{Balakumar Velayutham}
\email{vbalakumar2003@gmail.com}
\affiliation{%
  \institution{International Institute of Information Technology, Hyderabad}
  \city{Hyderabad}
  \country{India}
}

\author{Jashn Arora}
\email{arorajashn@google.com}
\affiliation{%
  \institution{Google DeepMind}
  \city{Bangalore}
  \country{India}
}

\author{Manish Jain}
\email{manishjn@google.com}
\affiliation{%
  \institution{Google DeepMind}
  \city{Bangalore}
  \country{India}
}

\author{Ponnurangam Kumaraguru}
\email{pk.guru@iiit.ac.in}
\affiliation{%
  \institution{International Institute of Information Technology, Hyderabad}
  \city{Hyderabad}
  \country{India}
}

\renewcommand{\shortauthors}{Arvindh Arun et al.}
%% No italics, no superscripts
%% Use footnote or author note to identify equal contribution and/or contact author info

\begin{abstract}
The detection of controversial content in political discussions on the Internet is a critical challenge in maintaining healthy digital discourse. Unlike much of the existing literature that relies on synthetically balanced data, our work preserves the natural distribution of controversial and non-controversial posts. This real-world imbalance highlights a core challenge that needs to be addressed for practical deployment. Our study re-evaluates well-established methods for detecting controversial content. We curate our own dataset focusing on the Indian political context that preserves the natural distribution of controversial content, with only 12.9\% of the posts in our dataset being controversial. This disparity reflects the true imbalance in real-world political discussions and highlights a critical limitation in the existing evaluation methods. Benchmarking on datasets that model data imbalance is vital for ensuring real-world applicability. Thus, in this work, (i) we release our dataset, with an emphasis on class imbalance, that focuses on the Indian political context, (ii) we evaluate existing methods from this domain on this dataset and demonstrate their limitations in the imbalanced setting, (iii) we introduce an intuitive metric to measure a model's robustness to class imbalance, (iv) we also incorporate ideas from the domain of Topological Data Analysis, specifically Persistent Homology, to curate features that provide richer representations of the data. Furthermore, we benchmark models trained with topological features against established baselines.
\end{abstract}





\maketitle

\section{Introduction}
\begin{figure*}[t]
    \centering
    \includegraphics[width=0.9\linewidth]{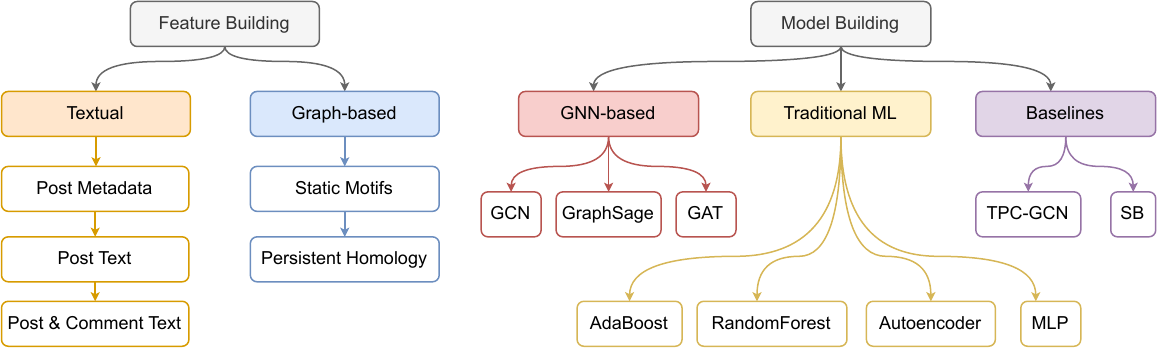}
    \caption{Taxonomy of the diverse features extracted and the methods used.}
    \label{fig:main}
\end{figure*}

In an era where digital platforms shape public discourse, Reddit stands out as a prominent forum for political discussions, often dubbed ``the front page of the Internet''. 2024 is a critical year in global politics, with more than half of the world's population participating in elections, including India's largest democratic exercise to date. As online platforms continue to shape political narratives \cite{politics1, politics2}, the role of ML systems in analyzing such discussions to generate valuable insights becomes paramount. Elections are deeply influenced by the dissemination of information and the framing of debates on digital platforms \cite{smelections}. ML systems enable the scalable analysis of these dynamics, uncovering patterns, detecting biases, and tracking the spread of misinformation, which are critical for understanding voter behavior and promoting informed electoral processes \cite{echochamber, smelections, healthy}.

Reddit's subreddits, particularly those focused on Indian political discourse, provide a rich dataset for examining the dynamics of online political engagement. These forums foster both constructive debates and contentious exchanges. Such politically active communities offer an opportunity to leverage ML systems to identify and analyze discussions where opposing views, including controversial ones, are actively debated. Detecting such discussions is crucial for uncovering emerging narratives, analyzing opinion bubbles, and assessing polarization in online communities \cite{echochamber, healthy}.

By curating a dataset from popular subreddits focused on Indian political discourse, we benchmark controversial content detection, typically marked by comparable participation from opposing views, on class-imbalanced data. This imbalance exists for several reasons like the echo chamber effect, where the users tend to gravitate towards communities that align with their existing beliefs, leading to a higher proportion of non-controversial discussions \cite{villa2021echo}; and the general social dynamics where people may avoid engaging in controversial discussions to maintain social harmony or avoid conflict \cite{bailey2024americans}. To emphasize the impact of significant class imbalance between controversial and non-controversial discussions that are inherent in political discourse, we re-evaluate well-established models in this domain. The primary contributions of our work are 1) introducing a new dataset and challenging existing evaluations to open a discussion on using class-imbalanced data to benchmark methods in this domain, 2) evaluating existing controversy detection methods on our dataset, 3) demonstrating the potential efficacy of using topological features for the task.

First, to establish a quantitative basis for analysing controversial discourse, we reintroduce a calibrated statistical definition of controversiality. Using this definition on our dataset reveals that truly controversial discussions, characterized by active debate and the presence of opposing viewpoints, are relatively rare compared to non-controversial discussions. The imbalance makes traditional evaluation metrics, such as overall accuracy and overall $F_1$ scores, less effective as they have a bias towards the majority class, leading to poor detection of controversial discussions and less reliable results. Addressing class imbalance is not a new problem, and various techniques like oversampling, undersampling, and the creation of ensemble models have been proposed in the past \cite{salunkhe2016classifier, tpc-gcn, dfe-gcn, smote}. We adopt a more generalized approach by curating features that capture the unique temporal evolution of each post, which provides more signal than static textural and structural features.

Next, we offer thorough evaluations of fusion-based methods in this domain, that leverage signals from both the textual content and the network structure of interactions on the content. We then benchmark Graph Neural Network (GNN) based approaches in this domain, which inherently learn the structural features in the data, providing insights into GNNs' effectiveness and limitations when tasked with performing on imbalanced real-world statistic datasets. We also introduce \textbf{Imbalance Impact Score} $(\mathcal{I})$, designed to quantify the performance disparity between balanced and imbalanced settings. Additionally, with inspirations from the Topological Data Analysis domain \cite{phomology}, specifically persistent homology, we curate more representative features and establish a new benchmark for assessing the performance of various methods in detecting controversial political content on imbalanced data.

The majority of previous works \cite{somethingsbrewing, wang2021mssf, multilingual, anger} in the domain of controversy detection primarily use artificially balanced datasets for testing, rendering their benchmarked results non-representative in the natural data distributions \cite{wikipedia, xposts, spanish}. We demonstrate the limitations of such benchmarks when scaled to an imbalanced dataset, with just 12.9\% of controversial posts, by reporting the Imbalance Impact Score. We propose new topology-based features with the aim of creating simple yet robust features for improving performance even in the imbalanced testing setting. Our work provides a comprehensive survey of baselines and features curated to understand and detect controversy in political discourse. Our results call for further research in this direction, emphasizing the need for scalable and robust approaches that can handle the complexities of real-world imbalanced datasets. To facilitate future studies and encourage reproducibility, we also release our dataset and code.\footnote{\url{https://drive.google.com/drive/folders/1L3nis3S-iiljLHVjvB5zaxxx-hkJufp6}} The overall taxonomy of the models and features we leverage is described in Figure \ref{fig:main}.

\begin{table*}[ht!]
\centering
\begin{tabular}{lccccrr}
\toprule
\textbf{Subreddit} & \textbf{Posts} & \textbf{Controversial} & \textbf{Non Controversial} & \textbf{Ratio} & \textbf{Users} & \textbf{Comments} \\
\midrule
\texttt{r/unitedstatesofindia}  & 4,832  & 549 (11.4\%) & 3810 (78.8\%) & $1:6.94$ & 31,713 & 326,034 \\
\texttt{r/India}                & 4,054  & 564 (13.9\%) & 3039 (74.9\%) & $1:5.39$ & 25,784 & 217,755 \\
\texttt{r/IndianModerate}        & 3,575  & 467 (13.1\%) & 2692 (75.3\%) & $1:5.76$ & 3,102 & 90,266  \\
\texttt{r/IndiaMeme}            & 2,856  & 159 (5.57\%) & 2555 (89.4\%)  & $1:16.1$ & 24,562 & 102,220 \\
\texttt{r/IndiaSpeaks}    & 2,150  & 196 (9.12\%) & 1826 (84.9\%)  & $1:9.32$ & 19,647 & 114,421 \\
\texttt{r/GeopoliticsIndia}      & 2,013  & \ 90 (4.47\%)  & 1822 (90.5\%)  & $1:20.2$ & 5,681 & 59,399  \\
\texttt{r/IndiaNews}             & \ 765    & \ 87 (11.4\%) & \ 606 (89.6\%)   & $1:6.97$ & 8,020 & 36,680 \\ \midrule
Total             & 20,245    & 2,112 (10.4\%) & 16,350 (80.8\%)   & $1:7.74$ & 79,867 & 946,775  \\
\bottomrule
\end{tabular}
\caption{Statistics of subreddits included in our dataset. Post counts presented here indicate posts after the comment count filter. Users in this table are unique Reddit users.}
\label{table:stats}
\end{table*}

\section{Relevance to Society}
The ability to detect and analyze controversial content has far-reaching implications for a diverse range of stakeholders engaged in understanding and shaping public discourse. News organizations can use these insights to identify emerging narratives and gauge public sentiment on critical issues. Political organizations can leverage this knowledge to better understand concerns and craft more responsive policies. Academic researchers studying political communication and social dynamics can benefit from more accurate controversy detection methods to advance their fields.

Furthermore, analyzing controversial discussions has important implications for social media platforms, especially those operating in culturally diverse settings like India. Enhanced controversy detection can enable platforms to adopt more nuanced content curation and user engagement strategies. This could lead to more sophisticated content moderation techniques, that preserve the diversity of public discourse while addressing the risks associated with highly polarized discussions. It can also be adapted for digital conflict resolution and consensus-building platforms, fostering more constructive online interactions.

While prior studies \cite{garimella2016quantifying,ortiz2020vocabulary,jang2016probabilistic} have investigated controversy detection on social networks, our work is the first to evaluate these methods under real-world class imbalance conditions. By reflecting the true distribution of controversial and non-controversial content, our approach provides a more accurate assessment of model performance in practical online environments. These findings highlight the need to re-calibrate evaluation paradigms to incorporate imbalanced datasets, ultimately enabling the development of more reliable and robust controversy detection systems capable of addressing the complexities of online political discourse.

\section{Related Work}

\textbf{Controversy Detection.}
Reddit's political landscape is characterized by diverse, topic-specific subreddit communities that foster a wide range of political discussions, often leading to controversies.\footnote{\url{https://en.wikipedia.org/wiki/Controversial_Reddit_communities}} Reddit's open-sourced version characterizes controversy based on high engagement metrics and, most importantly, a near-equal ratio of up-votes to down-votes, directly correlating to the existence of opposing views on a particular post.\footnote{\url{https://github.com/reddit-archive/reddit/blob/master/r2/r2/lib/db/_sorts.pyx}} Following are some previous studies that have analyzed these controversies using various approaches, including sentiment analysis, user interaction analysis, and content-based feature extraction.

\citet{somethingsbrewing} developed a feature-based approach to controversy detection by combining post-level features with comment tree features. Their method, however, lacked the utilization of more representative models like the GNNs. To address this gap, \citet{tpc-gcn} introduced a Topic-Post-Comment (TPC) graph that integrates both structural and textual features using GNNs.

\citet{benslimane2021controversy} constructed an undirected graph representing user interactions and used Graph Neural Networks for classification. Emphasizing structural aspects, \citet{coletto2017motif} proposed using network substructure counting (motifs) to identify local patterns of user interaction, significantly improving classification accuracy. Recent works have also explored the role of emotions in controversy detection. \citet{anger} analyzed the relationship between anger and controversy on Reddit, providing insights into the emotional dynamics of controversial discussions.

However, purely network-based or text-based methods can overlook subtle topological signals of controversy, such as cyclical interactions or multi-scale structures in conversation threads. We incorporate Topological Data Analysis (TDA), which is particularly well-suited to capture these phenomena by identifying loops and higher-dimensional ``holes'' in user-interaction graphs.

\textbf{Class Imbalance.}
The above-mentioned works have explored several techniques to artificially balance the classes, including upsampling, downsampling, and Synthetic Minority Over-sampling Technique (SMOTE) \cite{smote}. These methods aim to achieve a more balanced class distribution to evaluate their model performance, by sidestepping their evaluation in the real-world setting. Thus, despite the algorithmic advancements, their real-world generalizability has been limited

\cite{somethingsbrewing, jang2016probabilistic, ortiz2020vocabulary, multilingual, anger}. Our work addresses this limitation by evaluating controversy detection methods on a dataset that preserves the natural class imbalance found in real-world political discussions.

\section{Dataset}
\subsection{Collection}
To capture the dynamics of Indian political discourse online, we construct a dataset from Reddit, spanning over 10 months from 01-10-2023 to 20-07-2024, that maintains the original distribution of classes. We collected data from the 7 most popular (as per Reddit subreddit rankings)\footnote{\url{https://www.reddit.com/best/communities/1/}} subreddits focused on Indian politics. These subreddits collectively account for over 5 million active users, providing a comprehensive view of Indian political discussions online. Details about the initial dataset statistics are in Table \ref{table:stats}.

\begin{table}[h]
\centering
\begin{tabular}{c c}
\toprule
\textbf{Description} & \textbf{Stats} \\
\midrule
Total posts & 57,721 \\
Posts after comment count filter & 20,245 \\
\midrule
Posts after threshold filter & 18,462 \\
Controversial posts $(\mathrm{C})$ & 2,112 \\
Non-controversial posts $(\mathrm{NC})$ & 16,350 \\
Ratio of $\mathrm{C}$ to $\mathrm{NC}$ & 1 : 7.74 \\
\midrule
Total comments & 946,775\\
Median comments per post & 20 \\
Average nodes per post & 24.75 \\
Average edges per post & 38.22 \\
Average degree per post & 1.43 \\
\bottomrule
\end{tabular}
\caption{Filtered dataset statistics.}
\label{tab:stats2}
\end{table}

For \textit{r/India, r/IndiaSpeaks, r/UnitedStatesofIndia, r/IndiaMeme}, and \textit{r/IndiaNews}, we collected all posts tagged with the political flair, as added by the post authors or subreddit moderators. For \textit{r/GeopoliticsIndia} and \textit{r/IndianModerate}, we collected all posts without any filter, as the majority of the content in these subreddits is related to Indian politics. Each collected post includes rich metadata about the post itself and its associated comments.

\subsection{Controversy}

Detecting controversial posts requires first establishing a robust and context-specific definition of controversy. The metadata of all posts has an ``Upvote Ratio'' ($\mathrm{UR}$) field, which is defined as the fraction of upvotes on the total interactions (upvotes + downvotes) of the post. As shown by \citet{somethingsbrewing}, this is a fairly accurate proxy metric for the polarization observed in user perception of the post. Rather than labeling by the top and bottom quartiles after ranking and sorting by $\mathrm{UR}$ like \citet{somethingsbrewing}, we adopt a more generalized approach by directly utilizing the $\mathrm{UR}$ ranges.

\begin{figure}[t]
    \centering
    \includegraphics[width=8cm]{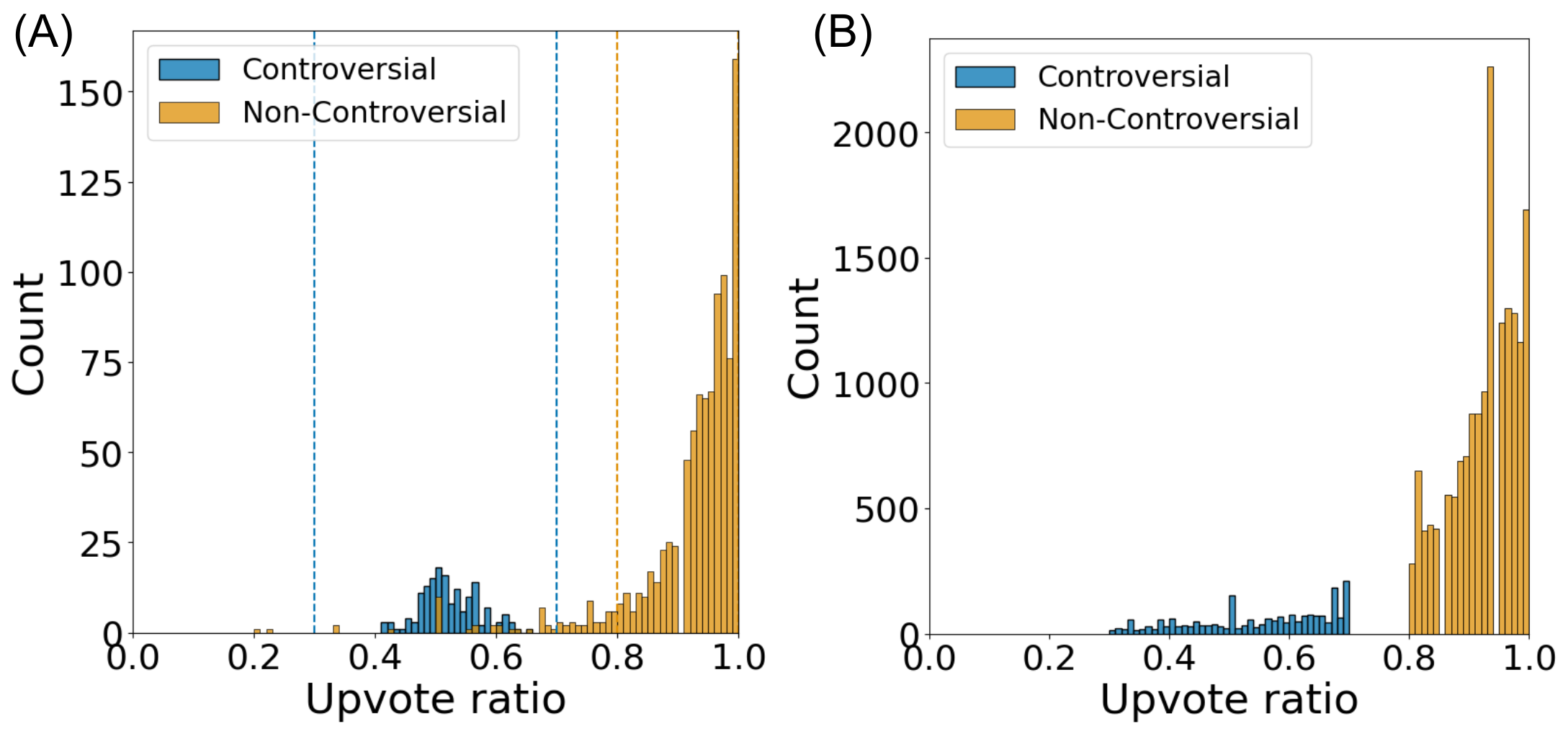}
    \caption{This density plot illustrates the distribution of $\mathrm{UR}$ for both controversial and non-controversial posts in our pilot study (A) and in our dataset (B). The plots reveal a region of separability between the two classes indicated by the vertical lines, which is used to derive a threshold for categorizing posts as controversial or non-controversial.}
    \label{fig:ur-dist}
\end{figure}

\textbf{Pilot Study.} To determine the range of $\mathrm{UR}$ in which controversial and non-controversial posts lie, we conduct a pilot study where we utilize Reddit's native feed filtering. Reddit provides various feed categories, including top, hot, and controversial posts. We categorize posts as follows:
\begin{itemize}
    \item Posts appearing in the ``controversial'' feed are labeled as controversial.
    \item Posts appearing in the ``top'' or ``hot'' feeds but not in the ``controversial'' feed are labeled as non-controversial.
\end{itemize}

For this study, we utilize posts from the last day of data collection, and using this categorization, we analyze the distribution of $\mathrm{UR}$ for both controversial and non-controversial posts. This analysis allows us to identify the characteristic $\mathrm{UR}$ ranges for each category and establish a quantitative basis for detecting controversial content. Based on the observations from the density distribution in Figure \ref{fig:ur-dist} (A), we define the following range of values:

\begin{itemize}
    \item[a)] $0.30 \leq \mathrm{UR} \leq 0.7$ : Controversial
    \item[b)] $0.80 \leq \mathrm{UR} \leq 1.00$ : Non-controversial
\end{itemize}

Using the above range of values to determine the controversial and non-controversial posts, we find that 98\% of posts were in agreement with the original pilot study. The density distribution on our dataset is shown in Figure \ref{fig:ur-dist} (B). This forms the basis of our labeling on the extracted dataset.

\begin{figure}[h]
    \centering
    \includegraphics[width=\linewidth]{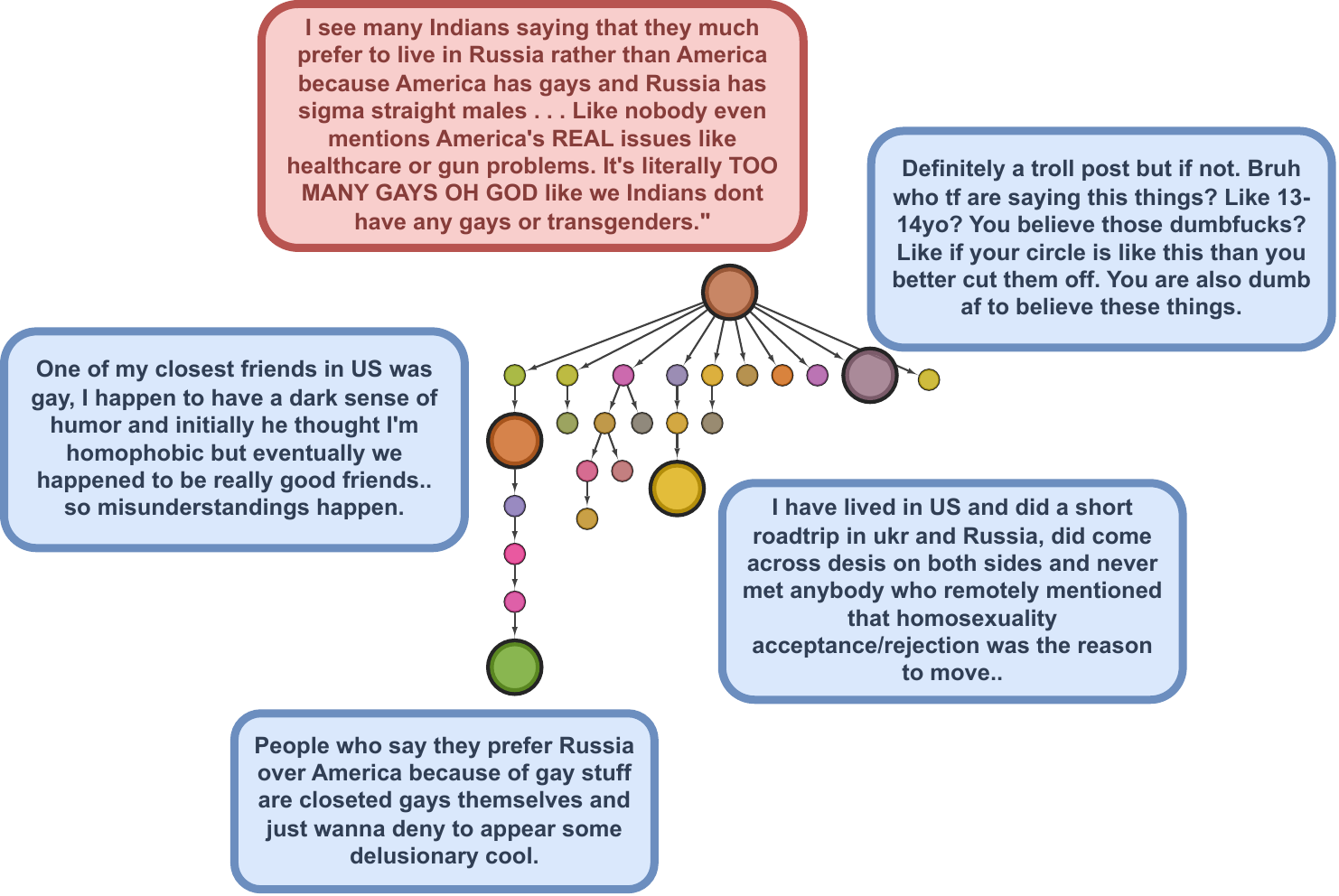}
    \caption{\textbf{\textcolor{red}{WARNING:}} The following figure contains potentially offensive language. The Post-Comment Tree of a controversial post (\#17i72r4) reveals deep branching and multiple levels of user interaction, highlighting the complexity and depth of engagement.}
    \label{fig:ct-ex}
\end{figure}

Applying this upvote ratio filter to the dataset, 1,783 out of 57,721 (3\%) posts are removed. To further enhance the dataset's quality, we eliminate posts with fewer than 5 comments. This refinement leaves us with a final count of 18,462 posts after applying the comment threshold filter. Overall statistics of the final filtered dataset are in Table \ref{tab:stats2}. An example of a sample controversial post's comment tree, where the nodes are posts and comments, is shown in Figure \ref{fig:ct-ex}.

\subsection{Graph Construction}
\begin{figure}[h]
    \centering
    \includegraphics[width=\linewidth]{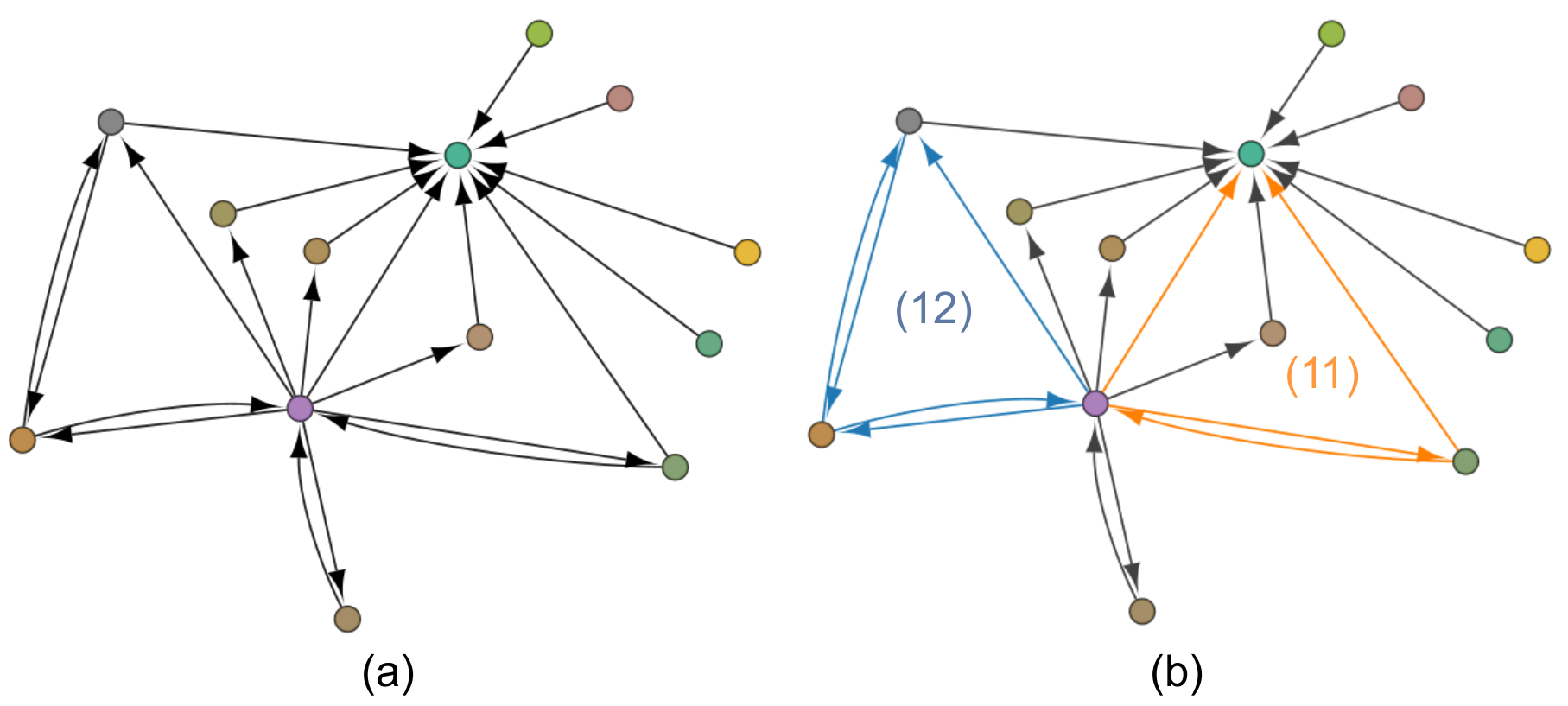}
    \caption{Subfigure (a) shows $G$ of a controversial post (\#17i72r4) revealing patterns of cyclic interactions, indicated by motifs in (b) where groups of users repeatedly interact with each other, often with contradicting viewpoints. In (b), the motifs highlighted in orange and blue correspond to the motif types (11) and (12) described in Figure \ref{fig:motifs}.}
    \label{fig:ui-ex}
\end{figure}

Previous studies \cite{somethingsbrewing, tpc-gcn} have shown that modeling user and post interactions as a graph can offer better signals for this task. To capture the intricacies of interactions between the users, we construct a weighted User-User interaction graph $(G)$ where the set of nodes are the users who commented on the post, including the author of the post. An edge exists between two nodes if they have replied to each other's comments at least once. An example is shown in Figure \ref{fig:ui-ex}.

\section{Feature Extraction}
Understanding and detecting controversial discussions requires a comprehensive feature set that captures multiple dimensions of discourse. To this end, we carefully selected features that address the structural, temporal, and content-based aspects of online interactions. The Python Reddit API Wrapper (PRAW)\footnote{\url{https://praw.readthedocs.io/en/stable/index.html}} provides a detailed set of features for the posts and users. We extract three major kinds of features from the data to capture the major modalities of signals. While PRAW provides static features of the post and its metadata, apart from that, we extract features capturing textual content to identify the subject of the discussion; static graph features capturing the interaction between users, and the temporal evolution of the interactions to fingerprint each post's own evolution. These features were chosen based on their demonstrated relevance in prior literature and their alignment with the multifaceted nature of controversial content \cite{somethingsbrewing, multilingual}.

\textbf{Post and User Interaction Features ($f_0$).} We curate a feature vector from the post metadata, specifically containing the number of comments, the number of users interacting with the post, the number of interactions between the users, and the average degree of interaction of users.

\textbf{Post Text Features $(f_1)$.} We start by extracting SBERT \cite{sbert} embeddings from the textual fields in the data. We use the pre-trained \texttt{all-mpnet-base-v2} model to generate 768-dimensional embeddings for two textual content levels. The first level, $f_1$, combines the post title and post text, capturing the essence of the initial post and providing a baseline representation of the topic. In the example from Figure \ref{fig:ct-ex}, this would combine ``Indians opinion on living in Russia or America.'' with ``I see many Indians saying that they much prefer to live in Russia rather than America because ...''. This feature set encodes the semantic content and contextual information of a post, enabling the model to represent the primary discussion topic effectively. By utilizing SBERT’s contextual embeddings, $(f_1)$ facilitates the extraction of patterns in the textual data that are critical for tasks such as identifying sentiment shifts or classifying controversial topics.

\textbf{Post + Comment Text Features $(f_2)$.} The second level, $f_2$, expands on this by incorporating the entire discussion, including all comments. This feature captures the entire discourse and the diverse perspectives introduced by commenters. Using insights from \citet{somethingsbrewing}, we apply mean-pooling to aggregate the individual comment representations to create a unified embedding. This allows us to capture the overall semantic content of the post and its discussion in a fixed-size vector, regardless of the length of the text or the number of comments.

\begin{figure}[h]
    \centering
    \includegraphics[width=7.5cm]{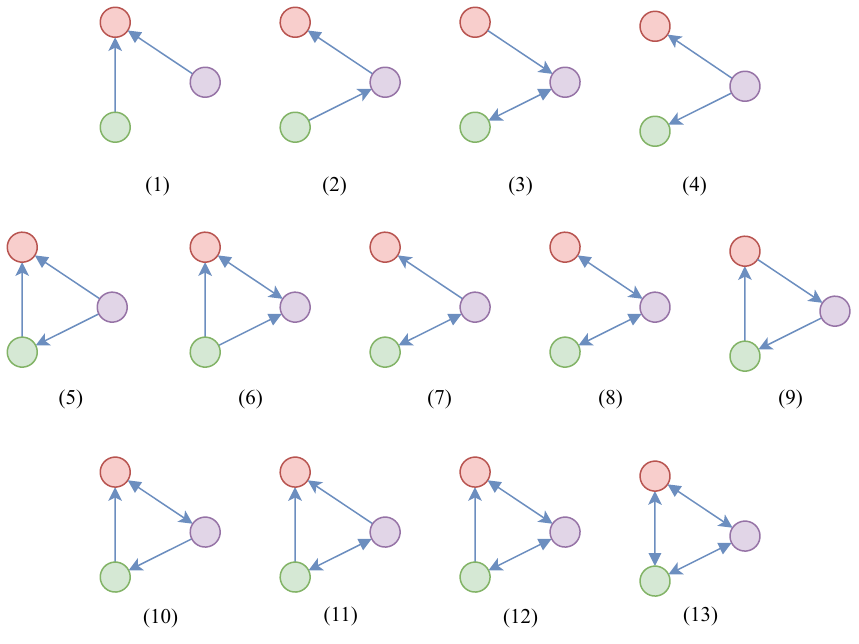}
    \caption{The 13 possible 3-motifs we count where each motif represents a different pattern of interactions among three users in a discussion. Counting these motifs provides insights into the interaction dynamics, such as agreement, disagreement, and the formation of echo chambers within the conversation.}
    \label{fig:motifs}
\end{figure}

\textbf{Static Graph Features $(f_3)$.} Next, we extract the 13 possible 3-motifs from $G$, the user-user interaction graph, as listed in Figure \ref{fig:motifs}. This simple yet informative feature captures repeating structural patterns in networks \cite{motifs}. We create a 13-dimensional vector, where each entry denotes the count of the corresponding substructure in the graph. These motifs provide insights into the complex interaction patterns within the discussion, potentially indicating the level of back-and-forth debate or the presence of echo chambers as observed in Figure \ref{fig:ui-ex}. For example, the presence of motif 11 can possibly indicate two users agreeing with a view and arguing against one user with an opposing view.

\begin{table*}[t]
\centering
\begin{tabular}{cccccc@{\hskip 0.5cm}ccc@{\hskip 0.5cm}ccc}
\toprule
 & & & & \textbf{AdaBoost} & & & \textbf{MLP} & & & \textbf{RF} &\\
\cmidrule(r){4-6} \cmidrule(r){7-9} \cmidrule(r){10-12}
\textbf{Training} & \textbf{Features} & & $F_c$(a) & $F_c$(c) & $\mathcal{I} (\uparrow)$ & $F_c$(a) & $F_c$(c) & $\mathcal{I} (\uparrow)$ & $F_c$(a) & $F_c$(c) & $\mathcal{I} (\uparrow)$ \\
\midrule
\multirow{5}{*}{(A)}
                        & $f_0$       & & 0.6727 & 0.2853 & 11.757  & 0.6722 & 0.2897 & 12.025  & 0.6291  & 0.2667  & 10.698  \\
                        & $f_1$       & & 0.6580 & 0.3256 & 14.303  & 0.6606 & 0.3171 & 13.752  & 0.6559 & 0.3421  & 15.397 \\
                        & $f_2$       & & 0.6704 & 0.3394 & 15.222  & 0.6925 & 0.3389 & 15.170  & 0.6822 & 0.3485 & 15.841 \\
                        & $f_2+f_3$      & & 0.7252 & 0.3802 & 18.060  & \underline{0.6988} & 0.3533 & 16.159  & 0.7264  & 0.3892  & 18.738  \\
                        & $f_2+f_3+f_4$      & & \underline{0.7287} & \underline{0.3839} & \textcolor{black}{\textbf{18.329}}  & \underline{0.6988} & \underline{0.3566} & \textcolor{black}{\textbf{16.392}}  & \underline{0.7523}  & \underline{0.3943}  & \textcolor{black}{\textbf{19.044}}  \\
\midrule
\multirow{5}{*}{(B)}
                        & $f_0$       & & 0.6701 & 0.2871 & 11.870  & 0.6713 & 0.2794 & 11.406  & 0.6261  & 0.2769  & 11.283  \\
                        & $f_1$       & & 0.6115 & 0.3309 & 14.557  & 0.6110 & 0.3910 & 18.634  & 0.5832 & 0.3759 & 17.378 \\
                        & $f_2$       & & 0.6569 & 0.3531 & 16.148  & 0.6809 & 0.3627 & 16.838  & 0.6081 & 0.3554 & 16.151 \\
                        & $f_2+f_3$      & & 0.6701 & 0.404 & 19.868  & 0.6707 & 0.3666 & 17.111  & 0.6635  & \underline{0.4145} & \textcolor{black}{\textbf{20.654}}  \\
                        & $f_2+f_3+f_4$      & & \underline{0.6945} & \underline{0.4142} & \textcolor{black}{\textbf{20.703}} & \underline{0.7391} & \underline{0.3971} & \textcolor{black}{\textbf{19.312}}  & \underline{0.6644} & 0.3910  & 18.876  \\
\midrule
\multirow{5}{*}{(C)}
                        & $f_0$       & & 0.0045 & 0.0045 & 0.002  & 0.0045 & 0.0045 & 0.002  & 0.0767  & 0.0668  & 0.507  \\
                        & $f_1$       & & 0.1963 & 0.1759 & 3.382  & 0.4302 & 0.3338 & 12.976  & 0.0769  & 0.0720  & 0.551  \\
                        & $f_2$       & & 0.1833 & 0.1657 & 2.984  & 0.3966 & 0.2935 & 10.440  & 0.0552 & 0.0521 & 0.287 \\
                        & $f_2+f_3$      & & 0.2921 & 0.2492 & 6.967  & 0.3966 & 0.3131 & 11.381  & 0.1038  & 0.0966  & 0.995  \\
                        & $f_2+f_3+f_4$     & & \underline{0.2943} & \underline{0.2557} & \textcolor{black}{\textbf{7.235}}  & \underline{0.4598} & \underline{0.3660} & \textcolor{black}{\textbf{15.250}}  & \underline{0.1220}  & \underline{0.1116}  & \textcolor{black}{\textbf{1.347}} \\
\bottomrule
\end{tabular}
\caption{Traditional Methods. For almost all of the traditional methods in all of the training settings, adding $f_4$ improves $\mathcal{I}$. (A) represents the balanced data scenario, (B) represents the scenario where the minority class is oversampled and the majority class is undersampled, (C) represents the imbalanced scenario. Best $\mathcal{I}$ for each model in each setting is highlighted in \textbf{bold}. \underline{Underlined} values represent the best $F_c$ for each model each setting.}
\label{tab:trad}
\end{table*}

\textbf{Dynamic Graph Features $(f_4)$.} To capture the evolving nature of discussions, we utilize techniques from Topological Data Analysis, particularly Persistent Homology \cite{phomology}. For instance, repeated back-and-forth exchanges among a small group of users can form loops in the user-interaction graph. Persistent Homology captures these loops, potentially reflecting intense disagreements that often characterize controversial discourse. It provides a robust theoretical framework for examining complex and dynamically structured data and is particularly well-suited for controversy detection for several reasons. First, it effectively captures the temporal evolution of discussions by analyzing topological features across multiple timescales, reflecting how conversations develop over time. Second, its robustness to noise \cite{phnoise} makes it ideal for managing the often chaotic nature of online discussions. Third, Persistent Homology is adept at capturing multi-scale interactions, ranging from individual exchanges to broader group dynamics, which are crucial for understanding controversy. 
Following established practices in topological data analysis \cite{phomology, phnoise}, we construct persistence diagrams using the Vietoris-Rips complex on $G$ and transform these into fixed-size feature vectors using Giotto TDA \cite{giotto-tda}.

We employ the Vietoris Rips Complex on $G$ to create a persistence diagram, from which we generate and flatten the persistent image into a feature vector using Giotto TDA \cite{giotto-tda}. This approach captures the evolution of interaction dynamics while ensuring computational tractability. More details are in the Appendix.

By incrementally building this rich feature set, we aim to create the most representative set of features for our downstream models. Each feature type captures different aspects of the political discourse, from the content and structure of the discussion to its evolution over time and the broader context of user interactions.

\begin{table*}[t]
\centering
\begin{tabular}{cccccccccccc}
\toprule
\textbf{Training} & \textbf{Variant} & & & \textbf{GCN} & & & \textbf{GAT} & & & \textbf{GSAGE} & \\
\cmidrule(r){4-6} \cmidrule(r){7-9} \cmidrule(r){10-12}
& & & $F_c$(a) & $F_c$(c) & $\mathcal{I} (\uparrow)$ & $F_c$(a) & $F_c$(c) & $\mathcal{I} (\uparrow)$ & $F_c$(a) & $F_c$(c) & $\mathcal{I} (\uparrow)$ \\
\midrule
\multirow{2}{*}{(A)}
                        & Base         & & 0.62 & 0.31 & \textcolor{black}{13.584} & 0.63 & 0.30 & \textcolor{black}{12.967} & 0.64 & \underline{0.34} & \textcolor{black}{15.243} \\
                        & Base + $f_4$ & & \underline{0.70} & \underline{0.34}   & \textcolor{black}{\textbf{15.251}} & \underline{0.69} & \underline{0.32} & \textcolor{black}{\textbf{14.082}} & \underline{0.69} & \underline{0.34} & \textcolor{black}{\textbf{15.249}} \\
\midrule
\multirow{2}{*}{(B)}
                        & Base  & & 0.51 & 0.29 & \textcolor{black}{11.695} & 0.55 & 0.29 & \textcolor{black}{12.373} & 0.52 & 0.32 & \textcolor{black}{13.621} \\
                        & Base + $f_4$ & & \underline{0.69} & \underline{0.31} & \textcolor{black}{\textbf{13.059}} & \underline{0.68} & \underline{0.30} & \textcolor{black}{\textbf{12.920}} & \underline{0.65} & \underline{0.33} & \textcolor{black}{\textbf{14.880}} \\
\midrule
\multirow{2}{*}{(C)}
                        & Base  & & \underline{0.61} & \underline{0.34} & \textcolor{black}{\textbf{14.888}} & 0.53 & \underline{0.34} & \textcolor{black}{14.596} & 0.49 & \underline{0.33} & \textcolor{black}{13.583} \\
                        & Base + $f_4$ & & 0.51 & 0.33 & \textcolor{black}{13.801} & \underline{0.58} & \underline{0.34} & \textcolor{black}{\textbf{14.987}} & \underline{0.51} & \underline{0.33} & \textcolor{black}{\textbf{13.723}} \\
\bottomrule
\end{tabular}
\caption{GNNs benefit from the addition of dynamic graph features $(f_4)$, but they still experience notable performance drops in imbalanced scenarios. (A) represents the balanced data scenario, (B) represents the scenario where the minority class is oversampled and the majority class is undersampled, (C) represents the imbalanced scenario. Base refers to the node features (embeddings of the post / comment content of that node) of a post graph. }
\label{tab:gnn}
\end{table*}

\begin{table}
\centering
\begin{tabular}{cccccc}
\toprule
\textbf{Training} & \textbf{Features} & & & \textbf{AutoEncoder}\\
\cmidrule(r){4-6}
& & & $F_c$(a) & $F_c$(c) & $\mathcal{I} (\uparrow)$\\
\midrule
\multirow{5}{*}{(C)}
                        & $f_0$     & & 0.30 & 0.25 & 7.125 \\
                        & $f_1$     & & 0.27 & 0.22 & 5.643 \\
                        & $f_2$     & & 0.28 & 0.24 & 6.451 \\
                        & $f_2+f_3$    & & 0.28 & 0.25 & 6.79 \\
                        & $f_2+f_3+f_4$    & & \underline{0.32} & \underline{0.28} & \textcolor{black}{\textbf{8.602}} \\
\bottomrule
\end{tabular}
\caption{Results for AutoEncoder. The addition of $f_4$ improves $\mathcal{I}$ in both balanced and imbalanced settings. (C) represents the imbalanced data scenario}
\label{tab:anon}
\end{table}

\section{Evaulation Framework}
\subsection{Evaluation setting}
We report the F1 scores of the controversial class $(F_c)$ and the Imbalance Impact Score $(\mathcal{I})$. The Imbalance Impact Score is particularly valuable as it quantifies a model's robustness to class imbalance, a critical requirement in real-world applications. We benchmark various combinations of features, models, and training settings by testing them in both balanced and imbalanced settings. In specific, we have three training scenarios,
\begin{enumerate}
    \item Training with balanced dataset by random undersampling of non-controversial posts \textbf{(A)}
    \item Training with balanced but upsampled dataset by random undersampling of non-controversial posts and oversampling of controversial posts. We incorporate this setting to maintain the data balance but also increase the number of data points as compared to the previous setting \textbf{(B)}
    \item Training with the original class distribution \textbf{(C)}
\end{enumerate}
and two testing scenarios,
\begin{enumerate}
    \item Testing on balanced dataset by random undersampling of non-controversial posts \textbf{(a)}
    \item Testing on the original class distribution \textbf{(c)}
\end{enumerate}

which leads to 6 different settings for each model. $F_c$(a) and $F_c$(c) correspond to testing on the balanced dataset and with the original class distribution, repsectively. 

\subsection{Imbalance Impact Score $(\mathcal{I})$}

The Imbalance Impact Score is a measure of a model's robustness to class imbalance. The intuition behind introducing this metric is simple: a good robust model, designed to be deployed in the real-world, should not only perform well when the classes are equally distributed but should also be equally effective in the data-imbalanced setting. The Imbalance Impact Score $(\mathcal{I})$ is defined as,
\begin{center}
    $\mathcal{I} = 100 \cdot (F_c\text{(a)} \times F_c\text{(c)}) \cdot (1 - |F_c\text{(a)} - F_c\text{(c)}|)$
\end{center}

where $F_c\text{(a)}$ is the F1-score of controversial class in the balanced setting and $F_c\text{(c)}$ is the F1-score of the controversial class when tested over the original distribution. $\mathcal{I}$ is a simple statistical measure that rewards high performance in both the testing settings and penalizes the performance drop between the settings to ensure that the model's performance is consistent. The score ranges from 0 to 100, with higher values indicating better performance consistency across both settings. When a model performs equally well on balanced and imbalanced data, $\mathcal{I}$ increases quadratically with the F1 score, rewarding consistency. Conversely, when a model's performance degrades significantly in the imbalanced setting, $\mathcal{I}$ decreases proportionally to penalize the model's lack of robustness. Maximum score $(\mathcal{I} = 100)$ is observed when the performance across settings is the same and the highest ($F_c\text{(a)} = F_c\text{(c)} = 1$), whereas the minimum score $(\mathcal{I} = 0)$ is observed when the difference in performance across settings is the maximum ($|F_c\text{(a)} - F_c\text{(c)}| = 1$).

\section{Experiments}
\subsection{Model Building}
To robustly study the impact of class imbalance, we employ a variety of commonly used models in this domain to classify controversial Reddit posts. Our approach includes both traditional classifiers and GNNs, each chosen for specific strengths that may contribute to effective classification in this domain. We run extensive Hyperparameter tuning for each setting using Optuna \cite{optuna}. More details are in the Appendix.

\begin{table}[h]
\centering
\begin{tabular}{ccccc}
\toprule
\textbf{Training} & \textbf{Baseline} & $F_c$(a) & $F_c$(c) & $\mathcal{I} (\uparrow)$\\
\midrule
\multirow{3}{*}{(A)}
                        & TPC-GCN       & 0.6623 & 0.2412 & 9.248 \\
                        & SB        & 0.7103 & 0.3818 & 18.211   \\
                        & SB + $f_4$ & \underline{0.7299} & \underline{0.3899} & \textcolor{black}{\textbf{18.783}}   \\
\midrule
\multirow{3}{*}{(B)}
                        & TPC-GCN     & -      & -      & - \\
                        & SB      &  0.7144      &  0.4251      & 21.583 \\
                        & SB + $f_4$  &  \underline{0.7414}      &  \underline{0.4471}      & \textcolor{black}{\textbf{23.393}} \\
\midrule
\multirow{3}{*}{(C)}
                        & TPC-GCN         & 0.3733 & 0.1699 & 5.052 \\
                        & SB          & 0.4332 & 0.3644 & 14.7 \\
                        & SB + $f_4$  & \underline{0.4501} & \underline{0.3786} & \textcolor{black}{\textbf{15.822}} \\
\bottomrule
\end{tabular}
\caption{Baselines. $SB$ performs the best out of all the methods tested. The addition of $f_4$ improves $\mathcal{I}$ for the baselines in both balanced and imbalanced settings. (A) represents the balanced data scenario, (B) represents the scenario where the minority class is oversampled and the majority class is undersampled, (C) represents the imbalanced scenario where TPC-GCN cannot be tested due to oversampling constraints.}
\label{tab:baselines}
\end{table}

\subsubsection{Traditional ML Models.}
We benchmark our evaluation with various models like AdaBoost, MLP, and RandomForest to learn the decision boundary between the controversial and non-controversial classes. We use Autoencoders \cite{chen2018autoencoder} to effectively model an anomaly detection setting where the Autoencoder is trained over the distribution of the majority class (non-controversial features) and tested on the outlier class (controversial features). 

\subsubsection{Graph Neural Networks.}
To capture inherent topological features in the graph, we benchmark diverse GNN architectures like the GCN \cite{kipf2016semi}, GAT \cite{velivckovic2017graph}, and GraphSAGE \cite{gsage}. We model controversy detection as a graph classification task on $G$ with initial node features as their text embeddings.

\subsubsection{Baselines.}
While there are more recent works in the domain, TPC-GCN \cite{tpc-gcn} and features from \citet{somethingsbrewing}, which we refer to as SB, still serve as the fundamental baselines. Recent works like \cite{dfe-gcn} offer incremental improvements over TPC-GCN and SB while incurring heavy computational costs and complex architecture designs. To establish a solid yet effective baseline, we focus only on these two established approaches.

\subsection{Results}
Overall, our results highlight the challenges posed by class imbalance in real-world controversy detection tasks. While the addition of dynamic graph features $(f_4)$ shows promise in improving model performance, including imbalanced scenarios, the drop in $F_c$ scores from balanced to imbalanced settings remains substantial across all models and feature combinations. Interestingly, topology-driven features often align with repeated inter-user exchanges, which intuitively signal heated debates. This interpretability supports the idea that cycles and other structural patterns in discussions can serve as crucial indicators of controversy.

As observed in Table \ref{tab:trad}, all traditional models (AdaBoost, MLP, and Random Forest) experience a substantial drop in $F_1$ score for controversial posts when moving from testing setting (A) to (C). This drop is consistent across all feature combinations and training settings. When we add the dynamic graph features $(f_4)$ to the feature vector, we see an improvement in performance, also in the imbalanced testing scenario. This is reflected in $\mathcal{I}$ being the highest when $f_4$ is added. As seen in Table \ref{tab:anon}, the addition of $f_4$ leads to improvement in $\mathcal{I}$ both in balanced and imbalanced testing.

Table \ref{tab:gnn} shows the performance of GNN models (GCN, GAT, and GraphSAGE) with and without $f_4$. To effectively combine the GNN embeddings and the $f_4$, we use sequentially stacked self-attention and cross-attention layers to dynamically allocate importance to features. The addition of $f_4$ generally leads to improved performance with some exceptions.

Table \ref{tab:baselines} shows that baselines like TPC-GCN and SB also struggle with class imbalance, experiencing significant drops in $F_c$ scores when moving from balanced to imbalanced testing scenarios. SB, being just a feature-based method, still outperforms fusion-based methods like TPC-GCN in our evaluations. Adding $f_4$ to SB gives the best $\mathcal{I}$ across all the models we evaluated.

The concerning trend of drop in $F_c$ scores in the results highlights the need for further research into robust methods for handling class imbalance in controversy detection on social media platforms.

\section{Conclusion}
We find that benchmarks in the domain of controversy detection are not representative of real-world imbalanced class statistics. Our findings highlight class imbalance as a critical challenge and defining characteristic in controversy detection tasks. By providing a real-world dataset, and by bridging the gap between theoretical models and practical applications, our work sets a new standard for controversy detection in social media analysis. Our rich dataset focuses on Indian politics and will serve as the new benchmark in this area of research. Our extensive evaluation of existing approaches not only highlights the need for more research in this direction but also lays the foundations for the evaluation of novel methods.

Future research directions include exploring the transferability of our approach to other social media platforms and problem domains. We also hope to investigate our approach's potential for early detection of controversial content. We believe our contributions will ultimately lead to more robust and reliable tools for analyzing and moderating online political discussions, thereby eventually assisting with healthier online discussions.

\noindent \textbf{Ethical Statement.} All data collected and analyzed was publicly available, and user-specific information was anonymized to protect user privacy. One of our work's main contributions is to address potential biases, especially those that arise from the class imbalance in controversial content. Our work reports only aggregated metrics to avoid singling out individuals or groups.

\section{Limitations}
While our study offers valuable insights into controversy detection in online political discourse, we acknowledge potential limitations. Our dataset, though extensive, is confined to Reddit discussions on Indian politics in English, potentially limiting generalizability to other platforms, languages, or political contexts. Additionally, while our topological features offer novel insights, they may not capture all nuances of the dataset.

\bibliographystyle{ACM-Reference-Format}

\bibliography{main}


\appendix

\begin{table*}[]
\centering
\begin{tabular}{cccc}
\toprule
\textbf{Model} & \textbf{Parameter} & \textbf{Range/Values} & \textbf{Search Type}\\
\midrule
\multirow{2}{*}{Adaboost} & \text{$\mathrm{n\_estimators}$} & $[10, 100]$ & int\\
                          & \text{$\mathrm{learning\_rate}$} & $[1e^{-3}, 1]$ & log \\
\midrule
\multirow{2}{*}{Random Forest} & \text{$\mathrm{n\_estimators}$} & $[10, 100]$ & int \\
                               & \text{$\mathrm{max\_features}$} & $\{\mathrm{sqrt}, \mathrm{log_2}\}$ & categorical \\
\midrule

\multirow{5}{*}{MLP} & \text{$\mathrm{hidden\_layer\_sizes}$} & $\{32, 64, 128\}$ & categorical\\
                     & \text{$\mathrm{learning\_rate\_mlp}$} & $\{\mathrm{invscaling}, \mathrm{adaptive}, \mathrm{constant}\}$ & categorical\\
                     & \text{$\mathrm{activation}$} & $\{\mathrm{identity}, \mathrm{logistic}, \mathrm{tanh}, \mathrm{relu}\}$ & categorical\\
                     & \text{$\mathrm{solver}$} & $\{\mathrm{lbfgs}, \mathrm{sgd}, \mathrm{adam}\}$ & categorical\\
                     & \text{$\mathrm{alpha}$} & $[1e^{-3}, 1]$ & log\\
\midrule

\multirow{4}{*}{GNNs} & \text{$\mathrm{learning\_rate}$} & $[5e^{-5}, 1e^{-1}]$ & log\\
                     & \text{$\mathrm{decay}$} & $[0, 1]$ & log\\
                     & \text{$\mathrm{hidden\_dim}$} & $\left\{\frac{emb\_dim}{8}, \frac{emb\_dim}{4}, \frac{emb\_dim}{2}\right\}$ & categorical\\
                     & \text{$\mathrm{pooling\_method}$} & $\{\mathrm{max}, \mathrm{sum}, \mathrm{mean}, \mathrm{set2set}, \mathrm{attention}, \mathrm{sort}\}$ & categorical\\
\midrule

\multirow{5}{*}{Autoencoder} & \text{$\mathrm{learning\_rate}$} & $[1e^{-5}, 1e^{-1}]$ & log \\
                             & \text{$\mathrm{loss}$} & $\{\mathrm{mse}\}$ & categorical \\
                             & \text{$\mathrm{epochs}$} & $[10, 100]$ & int \\
                             & \text{$\mathrm{threshold}$} & $[10, 100]$ & int \\
                             & \text{$\mathrm{encoding\_dim}$} & $\{128, 256, 512\}$ & categorical \\
\bottomrule
\end{tabular}
\caption{Optuna Search Ranges}
\label{tab:optuna}
\end{table*}

\begin{table*}[h]
\centering
\begin{tabular}{cccc}
\toprule
\textbf{Parameter} & \textbf{Range/Values} & \textbf{Search Type} \\
\midrule
Regularization Strength & $\{10^{-100}, 10^{-5}, 10^{-4}, 10^{-3}, 10^{-2}, 10^{-1}, 10^{0}, 10^{1}\}$ & categorical \\
Model Type & \{SVM, Logistic L1, Logistic L2, Logistic L1/L2\} & categorical \\
Feature Standardization & \{Yes, No\} & categorical\\
\bottomrule
\end{tabular}
\caption{SB Search Ranges}
\label{tab:sb}
\end{table*}

\section{Persistent Homology}
\label{sec:homology}
\subsection{Fundamental Concepts in Topology and Homology.}
Topology is concerned with the properties of spaces that are invariant under continuous transformations -- stretching and bending, but not tearing or attaching. Homology, in this context, quantifies the presence of $n$-dimensional holes within a topological space, providing a robust algebraic descriptor. The $k$-th homology group $H_k$ of a space quantifies $k$-dimensional holes, where $k = 0, 1, 2, \ldots$. For instance, $H_0$ represents connected components, $H_1$ captures loops, and $H_2$ reflects voids or trapped volumes.

\subsection{Persistent Homology: Theory and Computation.}
Persistent homology studies the evolution of homology groups across multiple scales. The analysis begins with a point cloud data $X$ and a proximity parameter $\epsilon$, constructing a sequence of nested subspaces $X_1 \subseteq X_2 \subseteq \ldots \subseteq X_n$ based on $\epsilon$.

\begin{itemize}
    \item \textbf{Filtration:} A filtration is a nested sequence of simplicial complexes $K_1 \subseteq K_2 \subseteq \ldots \subseteq K_n$, built over the dataset $X$ by varying the proximity parameter $\epsilon$.
    \item \textbf{Simplicial Complexes:} Each complex $K_i$ in the filtration corresponds to a different value of $\epsilon$, where the connections between data points are established if they are within $\epsilon$ distance of each other.
    \item \textbf{Homology Groups:} For each $K_i$, homology groups $H_k(K_i)$ are computed to detect $k$-dimensional holes.
\end{itemize}

The persistent homology can be represented through the persistence diagrams, where each point in the diagram represents a topological feature across the filtration values. The persistence of a feature is measured from its birth (when it appears) to its death (when it merges or disappears), formally given by:

\begin{equation*}
    \text{Persistence} = \text{Death}(\epsilon) - \text{Birth}(\epsilon)
\end{equation*}

\subsection{From Persistence Diagrams to Persistence Images.}
The conversion of persistence diagrams to persistence images allows for the application of machine learning algorithms by transforming topological data into a more usable form. Persistence diagrams, while informative, present challenges for direct application in standard machine learning models due to their set-like nature and variable size. Persistence images offer a solution by converting diagrams into fixed-size, vectorized representations.

A persistence image is a 2D histogram or image generated from a persistence diagram by placing a weighted Gaussian at each point in the diagram, where the weights are typically functions of the persistence (i.e., the lifetime of homological features). The resulting image provides a compact and informative representation, capturing the essence of the topological features encoded in the persistence diagram. Mathematically, for a point $(b,d)$ in a persistence diagram representing a feature born at time $b$ and dying at time $d$, the associated weight might be $(d-b)$, emphasizing more persistent features. The persistence image $I$ is then defined as:

\begin{equation*}
    I(x, y) = \sum_{(b, d) \in D} w(b, d) \cdot \exp\left(-\frac{(x-b)^2 + (y-d)^2}{2\sigma^2}\right)
\end{equation*}

where $I(x, y)$ represents the pixel value at $(x, y)$ coordinate in the 2D plane, $D$ denotes the diagram, $w(b,d)$ is the weight function, and $\sigma$ controls the spread of the Gaussian blurs.

\subsection{Featurization}
We extract a vector corresponding to each post, capturing its topological evolution over time. Feature vectors can be extracted from these images by flattening the matrix of pixel values into a vector, or by applying further feature extraction techniques such as principal component analysis (PCA) or convolutional neural networks (CNNs) to capture more nuanced aspects of the data structure. These feature vectors then serve as input to machine learning models, facilitating the integration of topological features into predictive analytics.

\section{Compute Infrastructure}
Experiments are conducted with Intel(R) Xeon(R) Gold 5317 CPU @ 3.00GHz and two NVIDIA NVIDIA RTX 5000 Ada with combined 64GB GPU memory. The operating system of the machine is Ubuntu 22.04.4 LTS. As for software versions, we use Python 3.11.0, Pytorch 2.2.1, and CUDA 12.3.0.

\section{Hyperparameters}

% \subsection{Optuna}
We use Optuna \cite{optuna} to search for the best set of hyperparameters. The search ranges of hyperparameters for Traditional Models, Autoencoder, and GNNs are expanded upon in Table \ref{tab:optuna}. We follow \citet{somethingsbrewing} for searching the best values for SB as reported in Table \ref{tab:sb}. We do not conduct hyperparameter searches for TPC-GCN but rather use the values reported in their paper \cite{tpc-gcn}.

\end{document}